\documentclass[10pt, twocolumn]{IEEEtran}
\usepackage{color}
\usepackage{cite}
\usepackage{graphicx}
\usepackage{subfig}
\usepackage{amsmath}
\usepackage{amsfonts}
\usepackage{amssymb}
\usepackage{algorithm}
\usepackage{algorithmic}
\usepackage{graphicx}
\usepackage{verbatim}

\newtheorem{example}{Example}

\newtheorem{stopping criterion}{Stopping Criterion}

\newcommand{\beq}{\begin{equation}}
\newcommand{\eeq}{\end{equation}}
\newcommand{\beqnn}{\begin{equation*}}
\newcommand{\eeqnn}{\end{equation*}}
\newcommand{\beqy}{\begin{eqnarray}}
\newcommand{\eeqy}{\end{eqnarray}}
\newcommand{\beqynn}{\begin{eqnarray*}}
\newcommand{\eeqynn}{\end{eqnarray*}}
\newcommand{\bit}{\begin{itemize}}
\newcommand{\eit}{\end{itemize}}
\newcommand{\ben}{\begin{enumerate}}
\newcommand{\een}{\end{enumerate}}
\newcommand{\bex}{\begin{example}}
\newcommand{\eex}{\end{example}}

\newcommand{\balg}[1]{\begin{algorithm} \caption{#1}}
\newcommand{\ealg}{\end{algorithm}}

\newcommand{\balgc}{\begin{algorithmic}[1]}
\newcommand{\ealgc}{\end{algorithmic}}

\newcommand{\bary}{\begin{array}}
\newcommand{\eary}{\end{array}}
\newcommand{\bmx}{\begin{bmatrix}}
\newcommand{\emx}{\end{bmatrix}}
\newcommand{\bsmx}{\left[\begin{smallmatrix}}
\newcommand{\esmx}{\end{smallmatrix}\right]}
\newcommand{\bmxc}[1]{\left[\begin{array}{@{}#1@{}}}
\newcommand{\emxc}{\end{array}\right]}
\newcommand{\bcn}{\begin{center}}
\newcommand{\ecn}{\end{center}}

\newcommand{\Rbb}{{\mathbb{R}}}
\newcommand{\Zbb}{{\mathbb{Z}}}

\newcommand{\Rnbn}{\Rbb^{n \times n}}

\newcommand{\Rmbm}{\Rbb^{m \times m}}

\newcommand{\Zn}{\Zbb^{n}}

\newcommand{\sLOB}{{\scriptscriptstyle \text{LB}}}
\newcommand{\sOB}{{\scriptscriptstyle \text{B}}}

\newcommand{\A}{\boldsymbol{A}}

\newcommand{\I}{\boldsymbol{I}}

\newcommand{\Q}{\boldsymbol{Q}}
\newcommand{\R}{\boldsymbol{R}}

\newcommand{\Z}{\boldsymbol{Z}}

\renewcommand{\l}{\boldsymbol{l}}

\renewcommand{\u}{\boldsymbol{u}}
\renewcommand{\v}{\boldsymbol{v}}

\newcommand{\x}{{\boldsymbol{x}}}
\newcommand{\y}{{\boldsymbol{y}}}
\newcommand{\z}{\boldsymbol{z}}
\newcommand{\0}{{\boldsymbol{0}}}

\newcommand{\br}{{\bar{r}}}

\newcommand{\bbQ}{{\bar{\Q}}}
\newcommand{\bbR}{{\bar{\R}}}

\newcommand{\ty}{{\tilde{y}}}

\newcommand{\tby}{{\tilde{\y}}}

\newcommand{\hbx}{{\hat{\x}}}

\begin{document}

\title{GfcLLL: A Greedy Selection Based Approach for Fixed-Complexity LLL Reduction}

\author{Jinming~Wen and~Xiao-Wen~Chang
\thanks{This work was supported by NSERC of Canada grant 217191-17 and postdoc research
fellowship from Fonds de recherche Nature et technologies.}
\thanks{Jinming~Wen is with the Department of Electrical and Computer Engineering, University of Alberta, Edmonton T6G 2V4, Canada (e-mail: jinming1@ualberta.ca).}
\thanks{X.-W. Chang is with the School of Computer Science, McGill University,
Montreal, QC H3A 0E9, Canada (e-mail: chang@cs.mcgill.ca).}
}

\maketitle

\begin{abstract}
The LLL lattice reduction has been widely used to decrease the bit error rate (BER) of the Babai point,
but its running time varies much from matrix to matrix.
To address this problem, some fixed-complexity LLL reductions (FCLLL) have been proposed.
In this paper, we propose two greedy selection based FCLLL algorithms: GfcLLL(1) and GfcLLL(2).
Simulations show that both of them give Babai points with lower BER
in similar or much shorter CPU time than existing ones.
\end{abstract}

\begin{IEEEkeywords}
Integer least squares  problem, fixed-complexity LLL reduction, success probability,
GfcLLL.
\end{IEEEkeywords}

\section{Introduction}
In MIMO detection and some other applications, we need to estimate an unknown parameter vector
$\hbx\in \mathbb{Z}^n$ from
\begin{align}
\label{e:model}
 \y=\A\hbx+\v, \quad \v \sim \mathcal{N}(\boldsymbol{0},\sigma^2 \I),
\end{align}
where $\y\in \mathbb{R}^m$ is an observation vector,  $\A\in\mathbb{R}^{m\times n}$ is a full column rank model matrix
and $\v\in \mathbb{R}^m$ is a noise vector.

A common method to estimate $\hbx$ is to solve the following ordinary integer least squares (ILS) problem:
\beq
\label{e:ILS}
\min_{\x\in\Zn}\|\y-\A\x\|_2^2,
\eeq
whose solution is the maximum likelihood estimator of $\hbx$.
Since \eqref{e:ILS} is NP-hard, for some real-time applications,
a suboptimal solution, which can be produced quickly, is computed instead of solving \eqref{e:ILS}.
One often used suboptimal solution is the ordinary Babai  point $\x^\sOB$,
produced by the Babai's nearest plane algorithm \cite{Bab86}.
It is shown in \cite{ChaWX13} that the LLL reduction algorithm \cite{LenLL82}
can always increase  the success probability of $\x^\sOB$ which is the probability of $\x^\sOB=\hbx$.

In communications, the parameter vector $\hbx$ is often subject to a box constraint (after some transformations), i.e.,
\beq \label{e:box}
\hbx \in {\cal B}:=\{ \x: \l \leq \x \leq \u, \ \x\in \Zn\}.
\eeq
In this situation, one can first use the LLL reduction to get the LLL-aided ordinary Babai point,
 then round it into the constraint box ${\cal B}$ to get an estimate of $\hbx$.

The LLL reduction is useful to improve the accuracy of the Babai points
for both unconstrained and box-constrained cases \cite{ChaWX13} \cite{WenC17}.
However, its running time
varies much from matrix to matrix even for a fixed dimension.
Moreover, it was shown in \cite{JalSM08} that in the MIMO context,
the worst-case computational cost of the LLL reduction for reducing
$\A$ is not even bounded by a function of $n$.
This may cause problems for real-time communications applications, where
limited and known run-time is essential \cite{VetPSH09}, from the implementation point of view.
To address this issue,  some  so called fixed-complexity LLL (FCLLL) reduction algorithms
have been proposed \cite{VetPSH09,LinMH13,WenZM14}.
For a given $\A$, an FCLLL algorithm is to get a reduced matrix of $\A$
that is close to the LLL reduced matrix of $\A$ in
a computational cost more or less fixed for matrices with the same dimensions.
For the FCLLL algorithms in \cite{VetPSH09,LinMH13,WenZM14},
the number of sweeps or the number of tests of the Lov\'{a}sz condition is fixed,
while for the new FCLLL algorithms to be proposed in this paper, the number of column permutations is fixed.
Note that an FCLLL algorithm may have different numbers of arithmetic operations for
different matrices with the same dimensions.
However, the difference  is small.
Moreover, there is an upper bound on the complexity in terms of number of arithmetic
operations, which is truly fixed for the same dimensions.

In this paper, we will propose a new approach for FCLLL reduction.
Unlike existing approaches, which use predefined traversal order for selecting
two consecutive columns for size reductions and permutation,
our new approach uses a traversal order based on a greedy selection strategy.
It is motivated by increasing the success probability of the Babai point.
Two greedy selection strategies are proposed for this purpose, leading to
two FCLLL algorithms GfcLLL(1) and GfcLLL(2), respectively.
The first strategy was originally proposed in \cite{ChaYZ05} for computing the full LLL reduction and the other one is new and more effective.
The greedy approach takes more data communication time to find the columns
to do size reductions and column interchanges than approaches with fixed traversal order.
However, simulations show that both GfcLLL(1) and GfcLLL(2) can
produce Babai points with lower bit error rate (BER)
than the FCLLL algorithms proposed in \cite{VetPSH09,LinMH13,WenZM14},
with similar or much less CPU time.

The rest of this paper is organized as follows.
In Section \ref{s:reduction}, we introduce the LLL and FCLLL reductions. 
In Section \ref{s:fcLLL}, we present our new algorithms.
In Section \ref{s:sim}, we do some simulations to show the effectiveness and
efficiency of  the new algorithms. Finally  we summarize this paper in Section \ref{s:sum}.

\section{Background}\label{s:reduction}

In this section, we first introduce the LLL reduction and the success probability of the Babai point
which is the motivation for our new algorithm, then we briefly review some recent FCLLL algorithms
which will be used for comparisons later.

Let $\A$ in \eqref{e:model} have the following QR factorization
\beq
\label{e:QR}
\A= [\underset{n}{\Q_1}, \underset{m-n}{\Q_2}]\bmx\R \\ \0 \emx,
\eeq
where $[\Q_1, \Q_2]\in \Rmbm$ is orthogonal and $\R\in \Rnbn$ is upper triangular.
Define $\tby=\Q_1^T\y$ and $\tilde{\v}=\Q_1^T\v$, then \eqref{e:model}  can be transformed to
\beq
\label{e:modelR}
\tby=\R\hat{\x}+\tilde{\v}, \quad \tilde{\v} \sim \mathcal{N}(\boldsymbol{0},\sigma^2 \I).
\eeq

The ordinary Babai  (integer) point $\x^\sOB\in \Zbb^n$
found by the Babai nearest plane algorithm  \cite{Bab86}
is defined as
\beq \label{e:ck}
\begin{split}
& c_n=\ty_n/r_{nn}, \quad x_n^\sOB=\lfloor c_n\rceil, \\
& c_{i}=(\ty_{i} \!-\! \sum_{j=i+1}^nr_{ij}x_j^\sOB)/r_{ii}, \,  x_i^\sOB=\lfloor c_i\rceil, \, i=n\!-\!1, \ldots, 1.
\end{split}
\eeq

The ordinary Babai   point $\x^\sOB\in \Zbb^n$ can be used as an estimator of $\hbx$,
and its success probability is (see \cite{ChaWX13})
\begin{align}
& P(\R):= \Pr(\x^\sOB=\hbx)=\prod_{i=1}^n \phi(r_{ii}) ,  \label{e:prob}  \\
& \phi(r_{ii})= \sqrt{\frac{2}{\pi}}\int_0^{|r_{ii}|/(2\sigma)}\exp(-\frac{1}{2}t^2)dt.  \label{e:phi}
\end{align}

With the QR factorization \eqref{e:QR}, the LLL reduction algorithm \cite{LenLL82}
reduces  $\R$ to $\bbR$ via
$
\bbQ^T \R \Z = \bbR,
$
where $\bbQ  \in \mathbb{R}^{n\times n}$ is orthogonal,
$\Z\in   \mathbb{Z}^{n\times n}$ is unimodular (i.e., $\det(\Z)=\pm1$)
and  $\bbR\in \mathbb{R}^{n\times n}$ is an upper triangular matrix  satisfying
\begin{align}
&|\br_{ik}|\leq\frac{1}{2}|\br_{ii}|, \quad i=1, 2, \ldots, k-1, \label{e:criteria1} \\
&\delta \br_{k-1,k-1}^2 \leq   \br_{k-1,k}^2+ \br_{kk}^2,\quad k=2, 3, \ldots, n, \label{e:criteria2}
\end{align}
where $\delta$ is a constant satisfying $1/4 < \delta \leq 1$.

Define
$
\bar{\y}=\bar{\Q}^T\tilde{\y}, \ \ \bar{\v}=\bar{\Q}^T\tilde{\v}, \ \ \hat{\z}=\Z^{-1}\hbx,
$
then \eqref{e:modelR}  can be transformed to
$\bar{\y}=\bbR\hat{\z}+\bar{\v}, \quad \bar{\v} \sim \mathcal{N}(\boldsymbol{0},\sigma^2 \I).$
Using \eqref{e:ck}, we can obtain its  Babai  point $\z^\sOB$.
Then we get the LLL-aided ordinary Babai  point $\x^\sLOB=\Z\z^\sOB$,
which can be used to estimate $\hbx$.
If $\hbx \in {\cal B}$ (see \eqref{e:box}), after obtaining $\x^\sLOB$, we round it to the nearest point in ${\cal B}$, leading to the box-constrained LLL-aided   Babai point.

The LLL algorithm in \cite{LenLL82} starts with column $2$ of $\R$
and ends with column $n$ of $\R$. When the reduction processes at column $k$,
it first performs size reductions on $r_{ik}$ for $i=k\!-\!1\!:\!-1\!:\!1$,
and then checks if \eqref{e:criteria2} holds.
If so, the column index increases by 1; otherwise
it permutes columns $k\!-\!1$ and $k$ of $\R$ and  the column index decreases by 1.
One does not know exactly how many iterations are required to finish the reduction process
(here the number of iteration means  the number of tests on \eqref{e:criteria2} \cite{LinMH13}).
The FCLLL algorithm in \cite{VetPSH09}, to be referred to as fcLLL, is a modification of the LLL algorithm.
It always goes from column 2 to column $n$ and never comes back in the process.
But it repeats the process $J$ times,  where $J$ is a fixed positive integer,
resulting in $L=J(n-1)$ iterations.

A modification of fcLLL, referred to as  EfcLLL, was given in \cite{LinMH13}.
EfcLLL does only size reductions on the super-diagonal entries of $\R$.
Like the LLL algorithm \cite{LenLL82}, both fcLLL and EfcLLL start the iterations from the second column
and finish at the last column of $\R$.
Different from this traversal strategy, the most recent FCLLL, referred to as EnfcLLL,
was proposed in \cite{WenZM14}. It uses a novel two-stage column traversal strategy.

\section{New FCLLL Reduction Algorithms}\label{s:fcLLL}

As explained in Section \ref{s:reduction}, both fcLLL and EfcLLL
start the iterations from the second column
and finish at the last column which may not be effective in increasing the success probability of
the Babai point in fixed time.
EnfcLLL  uses a different traversal strategy and can improve  the performance significantly.
But like the previous ones, its traversal order is still fixed in advance.
Let us use an extreme case to explain why a fixed order, which ignores
the particularity of a channel matrix,   may not work well sometimes.
Suppose we are allowed to do only one column permutation,
then it is obvious that the order-fixed algorithms are unlikely to produce the best result.

The idea of our approach is that at each step, we choose two consecutive columns
to do size reduction and permutation so that we get highest improvement of the success probability
of the Babai point.

Given an upper triangular matrix $\R$, suppose that for any specific $k$,
\eqref{e:criteria2} is not satisfied after $r_{k-1,k}$ is size reduced, i.e.
\beq \label{e:lovasz_not}
\delta r_{k-1,k-1}^2 > \left(r_{k-1,k}-\left \lfloor \frac{r_{k-1,k}}{r_{k-1,k-1}}\right \rceil r_{k-1,k-1}\right)^2 + r_{kk}^2.
\eeq
After the size reduction on $r_{k-1,k}$ (i.e., applying a unimodular matrix to $\R$ from right so that \eqref{e:criteria1} holds),
permutation of the two columns and triangularization, we obtain $\bbR$ which satisfies
\begin{align}
\bar{r}_{k-1,k}&=r_{k-1,k}- \lfloor r_{k-1,k}/r_{k-1,k-1} \rceil r_{k-1,k-1}, \nonumber \\
\bar{r}_{k-1,k-1}&=\sqrt{\bar{r}^2_{k-1,k}+r_{kk}^2},
\label{e:Rbar-entry}\\
|\bar{r}_{kk}|&= |r_{k-1,k-1}r_{kk}/\bar{r}_{k-1,k-1}|.\nonumber
\end{align}
Note that the above operations decrease  $|r_{k-1,k-1}|$ and  increase $|r_{kk}|$.
Then by \eqref{e:prob},
$$
\frac{P(\bbR)}{P(\R)} = \frac{\phi(\bar{r}_{k-1,k-1})\phi(\bar{r}_{kk})}{\phi(r_{k-1,k-1})\phi(r_{kk})}
:=T_k.
$$
In \cite{ChaWX13} it is proved that $T_k>1$.
Ideally we wish to find $k$ such that $T_k$ is the largest, then  perform size reduction, column permutation and triangularization.
However, computing $\phi(\zeta)$ (see \eqref{e:phi}) involves numerical integrations and is  expensive.
Instead we will look at other more efficient greedy strategies.

In the following, we propose two different greedy selection strategies
to choose two columns of $\R$ to do reduction at each step.
The first greedy selection strategy is to find
$$
j =  \arg\max_{k}\{T_k^{(1)}: T_k^{(1)} = \frac{|r_{k-1,k-1}|}{|\br_{k-1,k-1}|}, \, \eqref{e:lovasz_not} \mbox{ holds} \}.
$$
If  the above $j$ does not exist,  $\R$ is essentially LLL reduced as it can become LLL reduced
after performing size reductions.
Otherwise, we perform a size reduction on $r_{j-1,j}$, permute columns $j-1$ and $j$ of $\R$,  and triangularize $\R$ by a Givens rotation.
After that, we update  $T_j^{(1)}$, $T_{j-1}^{(1)}$ (if $j>1$) and $T_{j+1}^{(1)}$ (if $j<n$)
(note that other $T_j^{(1)}$'s are not changed), and start the next iteration.
This greedy selection strategy is to find a pair of columns which can reduce the larger one of the two diagonal elements most significantly
and it was  first proposed in \cite{ChaYZ05} for computing the LLL reduction in solving an ILS problem
for GPS applications. Later the same strategy was used in \cite{ZhaoLJD12} and \cite{ZhaQW12a}.
One problem with this strategy is $|r_{k-1,k-1}|/|\br_{k-1,k-1}|$ is invariant with respect
to   scaling of $\R(:,k-1\!:\!k)$,
but the success probability of the Babai point changes by scaling.

The second greedy selection strategy is to find
$$
j = \arg\max_{k}\{T_k^{(2)}: T_k^{(2)} = \frac{1}{|r_{kk}|} -  \frac{1}{|\br_{kk}|} , \,  \eqref{e:lovasz_not} \mbox{ holds} \}.
$$
Note that in the LLL reduction, after columns $k\!-\!1$ and $k$ are permutated, $|r_{kk}|$ will increase, and $1/|r_{kk}|$ will decrease.
This strategy is to find columns  $j-1$ and $j$ such that  $1/|r_{jj}|$ decreases most.
We can rewrite $T_k^{(2)}$ as
$$
T_k^{(2)} = \frac{|r_{k-1,k-1}|-|\br_{k-1,k-1}|}{|r_{k-1,k-1}r_{kk}|},
$$
which is a relative gap between $|r_{k-1,k-1}|$ and $|\br_{k-1,k-1}|$
(note that  the denominator is the determinant  of the lattice $\{\R(k\!-\!1\!:\!k,k\!-\!1\!:\!k) \x | \x\in \Zbb^2\}$).
Like $T_k$, $T_k^{(2)}$ is variant with respect to  scaling of $\R(:,k-1\!:\!k)$.
Our numerical tests indicate that the two columns found by maximizing $T_k^{(2)}$
are more likely to the same as those found by maximizing $T_k$ than those found
by  maximizing $T_k^{(1)}$.

For the sake of convenience, when \eqref{e:lovasz_not} does not hold,
we set $T_k^{(i)}=0$ for $i=1,2$. In our algorithms, we suppose the maximum number of column permutations (denoted by $N$) is given.
The description of our algorithms is as follows: 

\begin{algorithm}[h!]
\caption{GfcLLL($i$)}   \label{a:GfcLLL}
\begin{algorithmic}[1]
  \STATE compute the QR factorization \eqref{e:QR}, set $\mbox{swap}=0$, $\Z=\I_n$;
   \STATE compute $T_k^{(i)}$ for $k=2:n$;
   \FOR{$j=1:N$}
    \STATE  find $j$ such that $T_j^{(i)}=\max_{2\leq k \leq n}T_k^{(i)}$;
  \IF{$T_j^{(i)}>0$}
     \STATE perform size reduction on $r_{j-1,j}$ and update $\Z$;
     \STATE permute columns $j-1$ and $j$ of $\R$ and triangularize $\R$, and update $\Z$ and $\Q$;
     \STATE update $T_j^{(i)}$, $T_{j-1}^{(i)}$ if $j>1$, and $T_{j+1}^{(i)}$ if $j<n-1$;
     \ELSE
           \STATE break;
  \ENDIF
   \ENDFOR
\end{algorithmic}
\end{algorithm}
Like $J$ in \cite{VetPSH09} and \cite{LinMH13}, and $N_{max}$ in \cite{WenZM14},
$N$ depends on applications. By using sone techniques similar to that for showing the LLL algorithm is a polynomial time algorithm in \cite{LenLL82},
we can derive a complexity result for GfcLLL(i), which depends on $N$.
Then for any specific application which has a fixed complexity requirement,
we can find $N$.

\section{Simulations} \label{s:sim}
In this section, we compare the efficiency and effectiveness of  GfcLLL(i) with existing FCLLL algorithms.
As the box-constrained Babai points found by applying fcLLL and EfcLL are the same and fcLLL is slower than EfcLLL,
we do not compare GfcLLL(i) with fcLLL.
Two greedy LLL algorithms were proposed in \cite{WenM16}
and one is faster than the other one.
For comparison, we modified the faster one by fixing the number of column permutations
so that it became  an FCLLL algorithm, and refer it to as
GfcLLL-WM.
In the tests, we took the parameter $\delta=1$.
All of the tests were performed with \textsc{Matlab} 2016b on a desktop computer with Intel(R) Xeon(R) CPU E5-1603 v4 working at 2.80 GHz.

For a fixed dimension, a fixed type of QAM and a fixed $E_b/N_0$,
we randomly generated $10^3$ complex channel matrices $\A$
whose entries independently and identically follow the standard complex normal distribution.
For each generated matrix, we randomly generated $10^3$ complex signal vectors $\hbx$
and $10^3$ complex Gaussian noise vectors $\v$,
resulting in $10^6$ instances of complex linear models.
Each complex instance was then transformed to an instance of the real model  \eqref{e:model}.

To compare  GfcLLL(i)  with other FCLLL algorithms, we control the number of column permutations each algorithm performs so that they  have similar costs.
We first fix the number of sweeps $J$ for EfcLLL.
For any channel matrix, we record the number of column permutations performed by EfcLLL, which is denoted by $K$.
Then for the same channel matrix, we set the number of column permutations for EnfcLLL, GfcLLL-WM and GfcLLL(i) as $\lfloor0.35 K \rceil$, $\lfloor0.7K\rceil$ and $\lfloor0.7K\rceil$, respectively.
Our simulations indicate that the above choices usually make the CPU time taken
by our GfcLLL(i) less than those taken by other algorithms.

\begin{figure}[!htbp]
\centering
\includegraphics[width=3.0in]{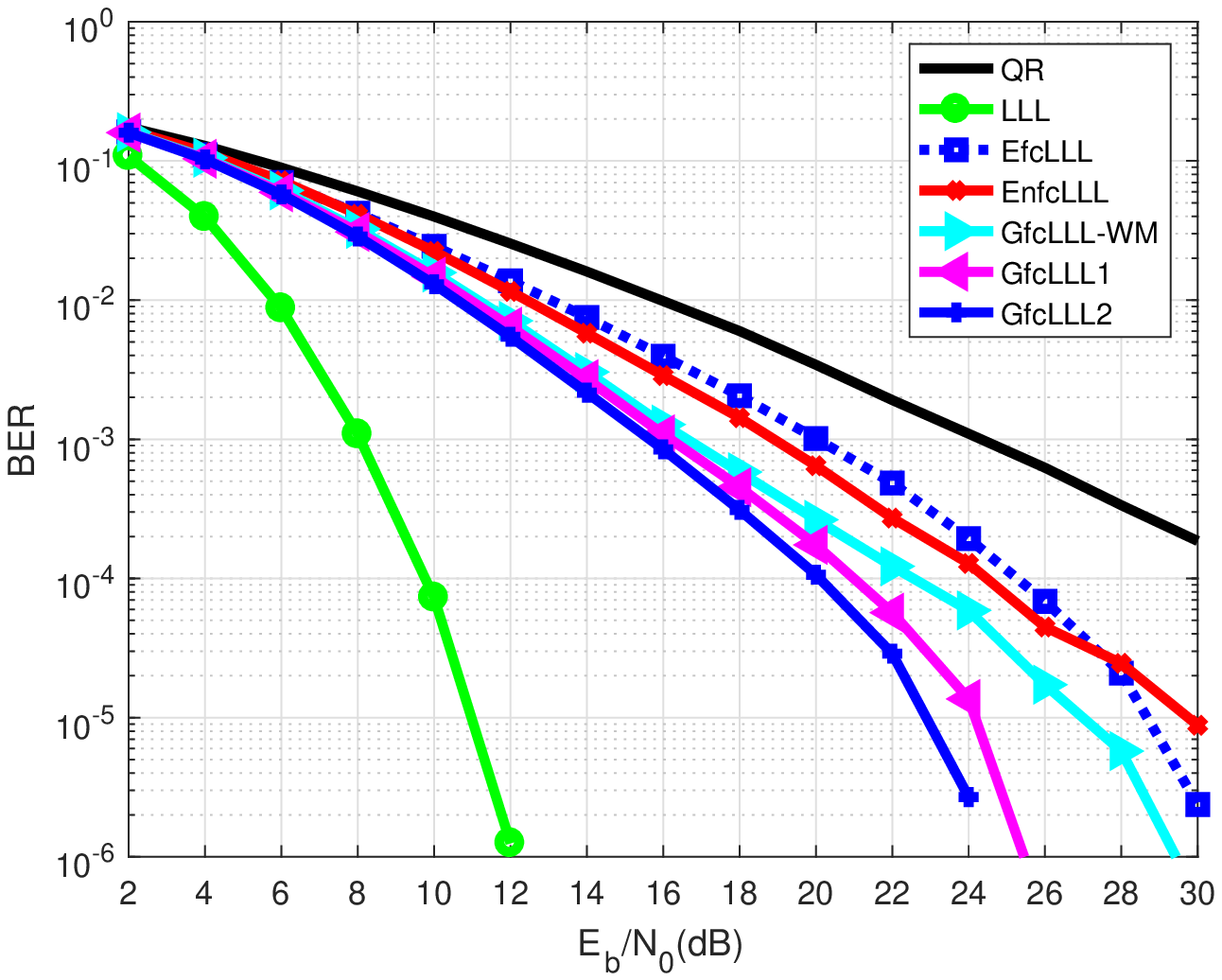}
\caption{BER versus $E_b/N_0=2\!:\!2\!:\!30$ for the $8\times8$ MIMO system with 4-QAM, $J=1$}
\label{fig:8D41YBER}
\end{figure}

\begin{figure}[!htbp]
\centering
\includegraphics[width=3.0in]{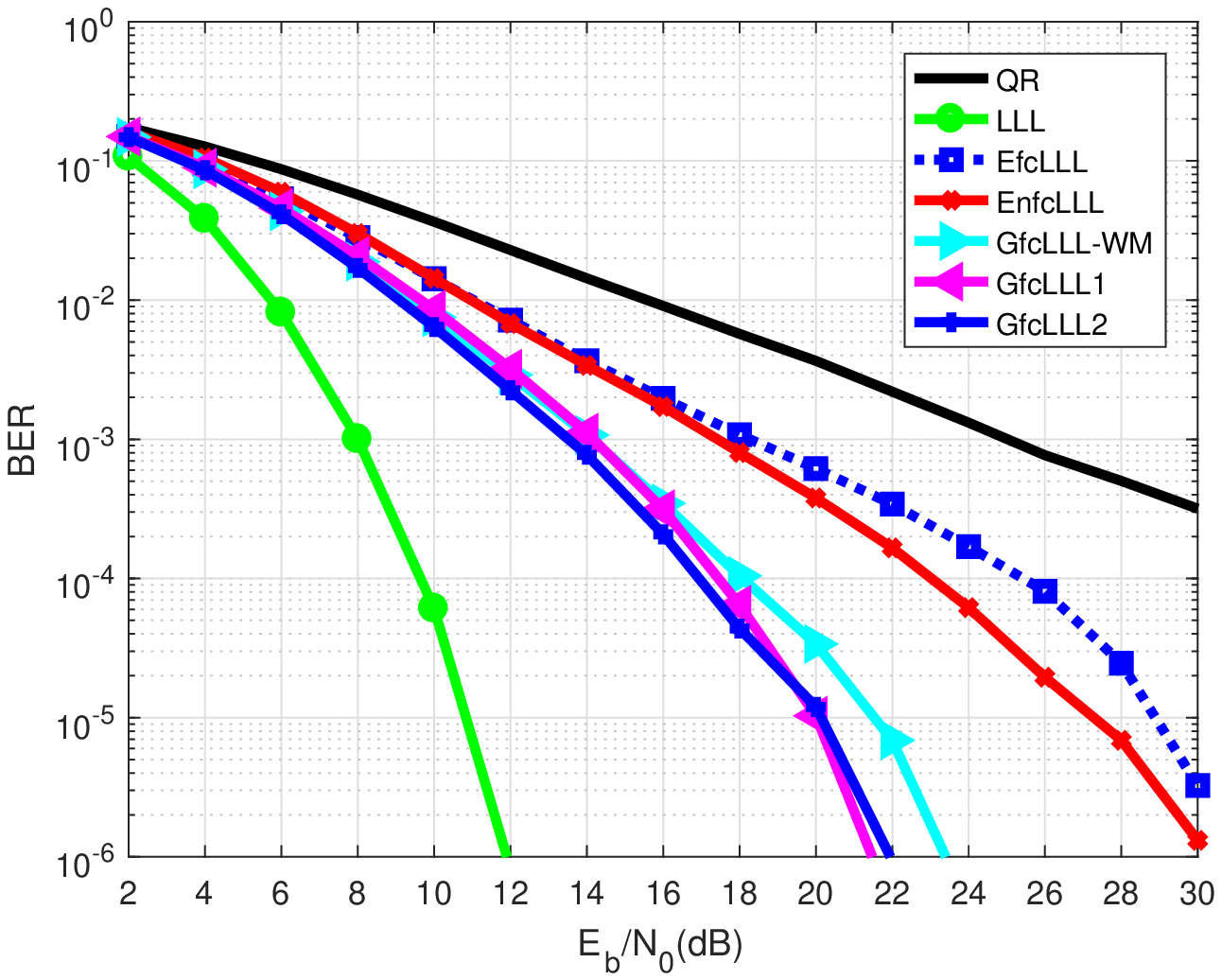}
\caption{BER versus $E_b/N_0=2\!:\!2\!:\!30$ for the $8\times8$ MIMO system with 4-QAM,  $J=2$}
\label{fig:8D42YBER}
\end{figure}

\begin{figure}[!htbp]
\centering
\includegraphics[width=3.0in]{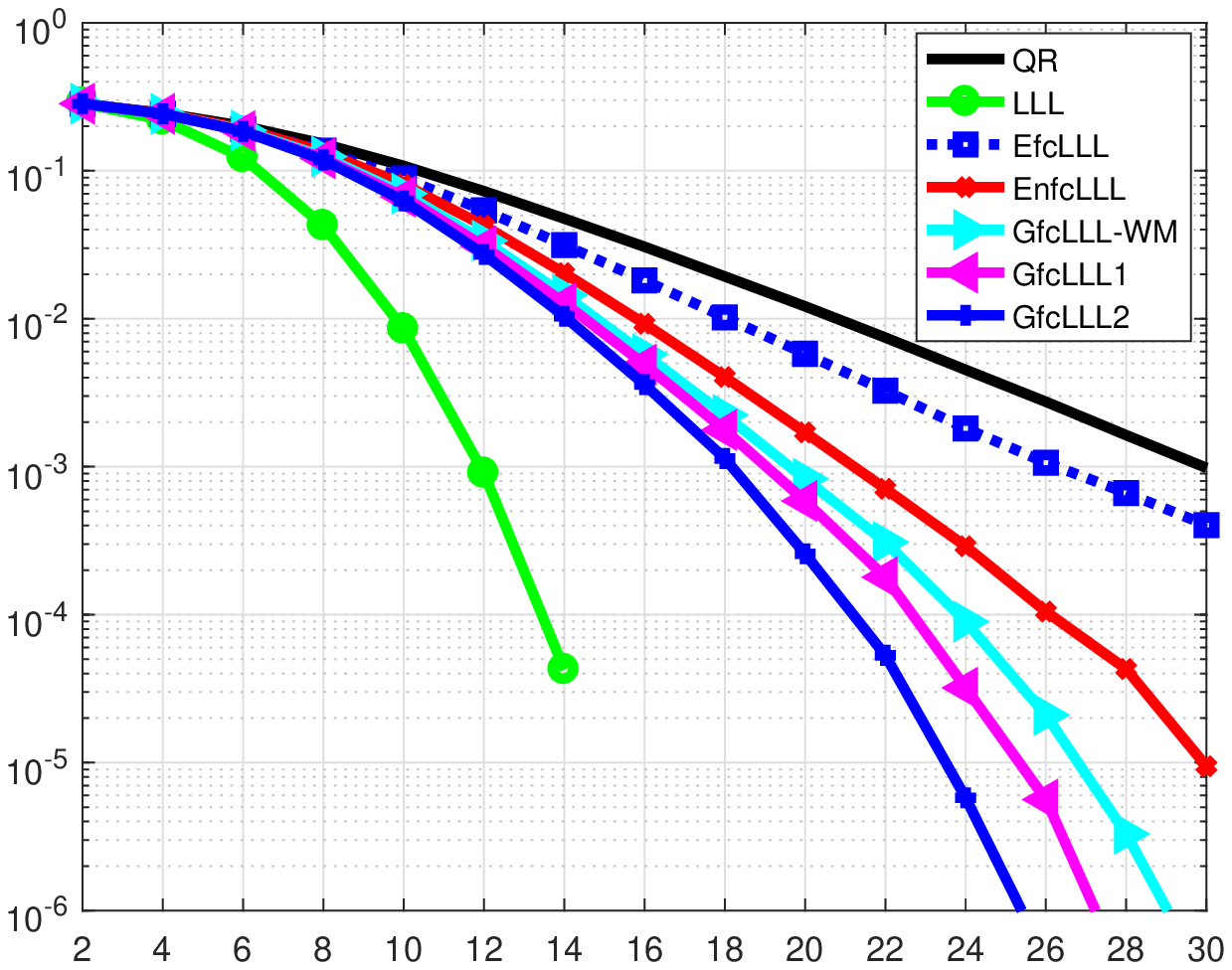}
\caption{BER versus $E_b/N_0=2\!:\!2\!:\!30$ for the $16\times16$ MIMO system with 16-QAM, $J=1$}
\label{fig:16D161YBER}
\end{figure}

\begin{figure}[!htbp]
\centering
\includegraphics[width=3.0in]{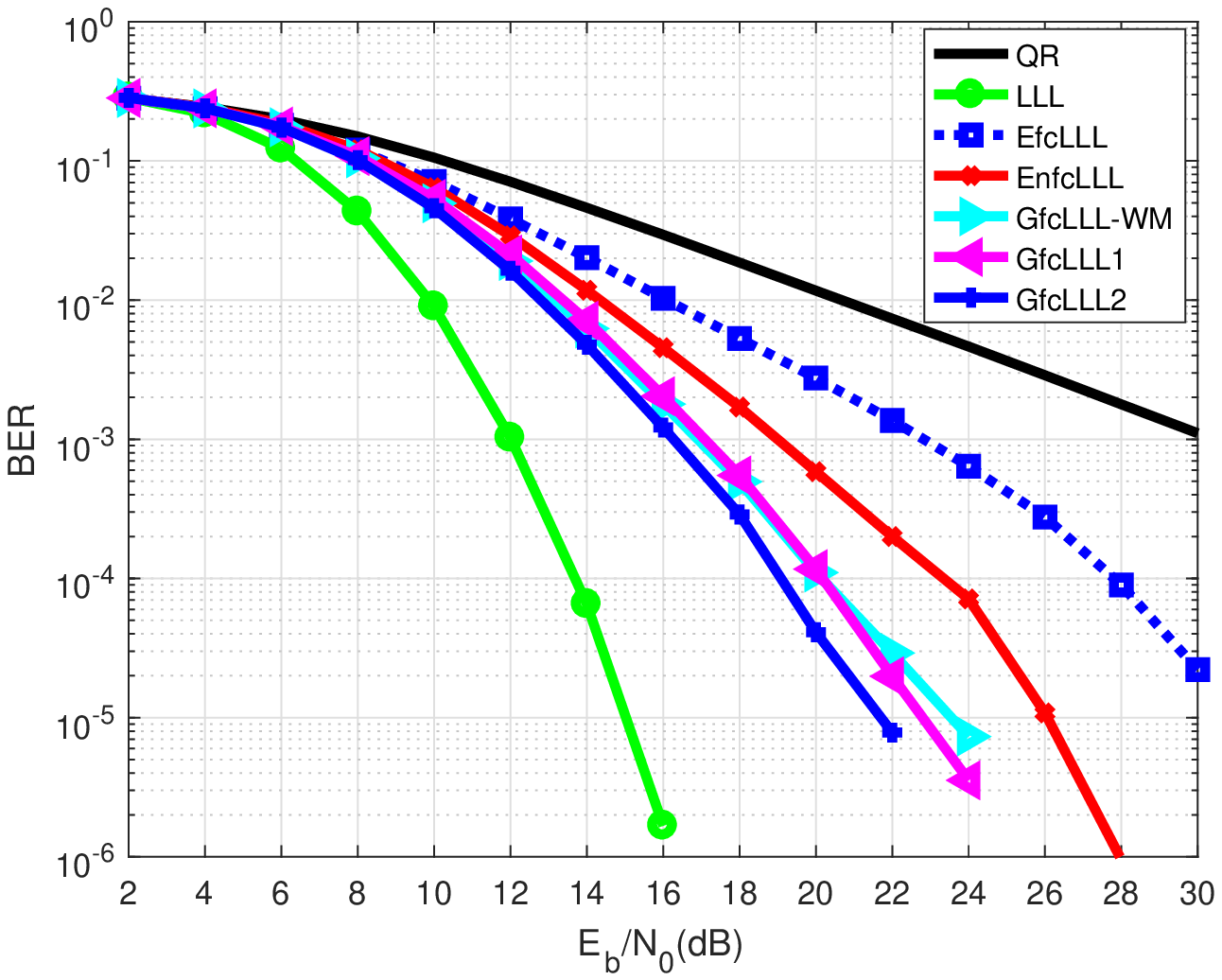}
\caption{BER versus $E_b/N_0=2\!:\!2\!:\!30$ for the $16\times16$ MIMO system with 16-QAM, $J=2$}
\label{fig:16D162YBER}
\end{figure}

Figures \ref{fig:8D41YBER} and \ref{fig:8D42YBER} show the average BER (over $10^6$ instances) versus
$E_b/N_0=2\!:\!2\!:\!30$ for the $8\times8$ complex
MIMO systems with 4-QAM for $J=1$ and $J=2$, respectively.
Similarly, Figures \ref{fig:16D161YBER} and \ref{fig:16D162YBER} show the corresponding results
for the $16\times16$ complex MIMO systems with 16-QAM for $J=1$ and $J=2$, respectively.

Tables \ref{tb:8D4YCpu} and  \ref{tb:16D16YCpu} respectively display the total CPU time of computing reductions for the $1000$  $8\times 8$ and $16\times 16$ complex channel matrices with  $J=1,2$.

\begin{table}[h!]
\scriptsize
\caption{Total CPU time for 1000 $8\times8$ channel matrices}
\centering
\begin{tabular}{|c||c|c|c|c|c|c|c|c|c|}
\hline
$J$ & LLL  & EfcLLL & EnfcLLL &GfcLLL-WM &GfcLLL1 &GfcLLL2 \\ \hline
$ 1$&    2.5333 &    0.3124  &  0.2783  &  0.2473  &  0.2392 &   0.2380 \\ \hline
$ 2$&   2.7438  &  0.5735   & 0.4315 &   0.4214 &   0.4061  &  0.4175 \\ \hline
\end{tabular}
\label{tb:8D4YCpu}
\end{table}

\begin{table}[h!]
\scriptsize
\caption{Total CPU time for 1000 $16\times16$ channel matrices}
\centering
\begin{tabular}{|c||c|c|c|c|c|c|c|c|c|}
\hline
$J$ & LLL  & EfcLLL & EnfcLLL &GfcLLL-WM &GfcLLL1 &GfcLLL2 \\ \hline
$1$ &   9.4890  &  0.6705  &  0.6488  &  0.5325  &  0.5058   & 0.5257  \\ \hline
$2$&   9.1026  &  1.1286   & 1.0107 &   0.8390  &  0.8135  &  0.8204 \\ \hline
\end{tabular}
\label{tb:16D16YCpu}
\end{table}

From Figures \ref{fig:8D41YBER}--\ref{fig:16D162YBER} and Tables \ref{tb:8D4YCpu} and \ref{tb:16D16YCpu},
we can see that the box-constrained Babai points  aided by  GfcLLL(i) ($i=1,2$)    have lower BER than those aided by existing FCLLL reductions,
while the former  cost less than the latter.
Thus our proposed GfcLLL(i) ($i=1,2$)  are more efficient and effective.
We also observe that GfcLLL(2) gives  better performance than GfcLLL(1)  with similar cost.
The simulation results also show that GfcLLL(i) can decrease the BER of the box-constrained Babai points
significantly by performing a small number of permutations after the QR factorization.

In  the above tests, EnfcLLL performed about a half number of permutations performed by GfcLLL(i).
In our simulations we found that if we set the  same number of permutations  for both EnfcLLL and GfcLLL(i),
the former is still worse than the latter in decreasing the BER of the  box-constrained Babai points
(note that in this case, the CPU time used by EnfcLLL is much higher than those by GfcLLL(i)).

\section{Summary} \label{s:sum}
We have proposed two greedy selection based FCLLL algorithms: GfcLLL(1) and GfcLLL(2).
Simulations showed that both result in the box-constrained Babai points with lower BER
in  shorter CPU time than existing FCLLL algorithms and GfcLLL(2) is more effective than GfcLLL(1).

\bibliographystyle{IEEEtran}
\bibliography{ref}
\end{document}